\documentclass[conference,10pt]{IEEEtran}
\usepackage[cmex10]{amsmath}
\usepackage{eqparbox}
\usepackage[usenames, dvipsnames]{color}
\usepackage{mathtools,amssymb,bm,mathabx,adjustbox}
\usepackage{amstext}
\usepackage{amssymb,amsfonts}
\usepackage{graphicx}
\usepackage{color}
\usepackage{booktabs}
\usepackage{longtable}
\usepackage{multicol}
\usepackage{algorithmicx}
\usepackage[ruled]{algorithm}
\usepackage{algpseudocode}
\usepackage{algpascal}
\usepackage{algc}
\usepackage{amsfonts}
\usepackage{dsfont}
\usepackage{array}
\usepackage[most]{tcolorbox}
\usepackage{cases}
\usepackage{pgfplots}
\usepackage[labelfont=bf,font=footnotesize]{caption}
\usepackage{subfigmat}
\usepackage{xr-hyper}
\usepackage{amsthm}
\DeclareCaptionLabelSeparator{periodspace}{.~}
\usepackage{float}
\usepackage{cite}
\theoremstyle{remark}

\newcommand\ASTART{\bigskip\noindent\begin{minipage}[b]{0.5\linewidth}}
	
	\newcommand\AENDSKIP{\end{minipage}\bigskip}
\newcommand\AEND{\end{minipage}}
\ifCLASSOPTIONcaptionsoff
\usepackage[nomarkers]{endfloat}
\let\MYoriglatexcaption\caption
\renewcommand{\caption}[2][\relax]{\MYoriglatexcaption[#2]{#2}}
\fi
\theoremstyle{plain}
\newtheorem{thm}{\textbf{Theorem}}
\newtheorem{lem}{\textbf{Lemma}}

\theoremstyle{definition}

\newcommand*{\rom}[1]{\expandafter\@slowromancap\romannumeral #1@}

\newcommand{\RN}[1]{%
\textup{\uppercase\expandafter{\romannumeral#1}}%
}

\usepackage{standalone}
\graphicspath{{./Figures/}}
\usepackage{times}
\usepackage[T1]{fontenc}
\usepackage[english]{babel}
\usepackage{graphicx}
\usepackage{ amsfonts, dsfont,amssymb, graphicx, array, tabularx, booktabs,multicol,nccmath}
\usepackage{amsthm}
\usepackage{epsf,epsfig}
\usepackage{setspace}
\usepackage{subfigure}
\usepackage{multirow}
\usepackage{url}
\usepackage{color}
\usepackage[nolist,printonlyused]{acronym}      % Acronym
\usepackage{comment}
\usepackage{cite}
\usepackage{bm}
\usepackage{tikz}
\usetikzlibrary{decorations.pathreplacing}

\usepackage{mathtools}

\usepackage{blindtext}

\usepackage{stfloats}

\newcommand{\mx}[1]{\mathbf{#1}}
\newcommand{\bs}[1]{\boldsymbol{#1}}

\definecolor{amber}{rgb}{1.0, 0.49, 0.0}
\definecolor{ao}{rgb}{0.0, 0.5, 0.0}

\def\R2#1{\textcolor{black}{#1}}
\def\R3#1{\textcolor{black}{#1}}

\renewcommand{\triangleq}{\mathbin{\setstackgap{S}{0pt}\stackMath\Shortstack{\smalltriangleup\\ =}}}
\usepackage[usestackEOL]{stackengine} %Needed to avoid tiny triangle in \triangleq
\usepackage[utf8]{inputenc} 
\usepackage{url}
\usepackage{ifthen}
\usepackage{cite}
\IEEEoverridecommandlockouts
\usepackage{cite}
\usepackage{amsmath,amssymb,amsfonts}
\usepackage{graphicx}
\usepackage{textcomp}
\usepackage{xcolor}
\def\BibTeX{{\rm B\kern-.05em{\sc i\kern-.025em b}\kern-.08em
    T\kern-.1667em\lower.7ex\hbox{E}\kern-.125emX}}

\makeatletter
\newcommand{\linebreakand}{%
  \end{@IEEEauthorhalign}
  \hfill\mbox{}\par
  \mbox{}\hfill\begin{@IEEEauthorhalign}
}
\makeatother
    \IEEEoverridecommandlockouts
\usepackage{cite}
\usepackage{graphicx}
\usepackage{textcomp}
\usepackage{xcolor}
\def\BibTeX{{\rm B\kern-.05em{\sc i\kern-.025em b}\kern-.08em
    T\kern-.1667em\lower.7ex\hbox{E}\kern-.125emX}}
\begin{document}

\title{Improved Downlink Channel Estimation in Time-Varying FDD Massive MIMO Systems\\
\thanks{This work is partly funded by Digital Futures. G. Fodor was also supported by the Swedish Strategic Research (SSF)
grant for the FUS21-0004 SAICOM project and the 6G-Multiband Wireless and Optical Signalling for Integrated Communications, Sensing and Localization (6G-MUSICAL) EU Horizon 2023 project,
funded by the EU, Project ID: 101139176. }
}
\author{\IEEEauthorblockN{1\textsuperscript{st} Sajad ~Daei}
\IEEEauthorblockA{\textit{School of Electrical Engineering and Computer Science} \\
\textit{KTH Royal Institute of Technology, Stockholm, Sweden}\\
Stockholm, Sweden,\\
Email: sajado@kth.se}
\and
\IEEEauthorblockN{2\textsuperscript{rd} Mikael Skoglund}
\IEEEauthorblockA{\textit{School of Electrical Engineering and Computer Science} \\
\textit{KTH Royal Institute of Technology, Stockholm, Sweden}\\
Stockholm, Sweden,\\
Email: skoglund@kth.se}
%\and
\linebreakand 
\IEEEauthorblockN{3\textsuperscript{nd} Gabor~Fodor}
\IEEEauthorblockA{\textit{School of Electrical Engineering and Computer Science} \\
\textit{KTH Royal Institute of Technology, Stockholm, Sweden}\\
\textit{ Ericsson Research, Stockholm, Sweden}\\
Stockholm, Sweden,\\
 Email: gaborf@kth.se}

}
% \and
% \and
% \and
% \and
% \and

\maketitle

\begin{abstract}
In this work, we address the challenge of accurately obtaining channel state information at the transmitter (CSIT) for frequency division duplexing (FDD) multiple input multiple output systems. Although CSIT is vital for maximizing spatial multiplexing gains, traditional CSIT estimation methods often suffer from impracticality due to the substantial training and feedback overhead they require. To address this challenge, we leverage two sources of prior information simultaneously: the presence of limited local scatterers at the base station (BS) and the time-varying characteristics of the channel. The former results in a redundant angular sparsity of users' channels exceeding the spatial dimension (i.e., the number of BS antennas), while the latter provides a prior non-uniform distribution in the angular domain. We propose a weighted optimization framework that simultaneously reflects both of these features. The optimal weights are then obtained by minimizing the expected recovery error of the optimization problem. This establishes an analytical closed-form relationship between the optimal weights and the angular domain characteristics. Numerical experiments verify the effectiveness of our proposed approach in reducing  the recovery error and consequently resulting in decreased training and feedback overhead.
\end{abstract}

\begin{IEEEkeywords}
Channel estimation, frequency division duplexing, multiple input multiple output, sparse representation. 
%\gf{\it GF: sparse representation is a keyword, but does not appear either in the title or in the abstract.}
\end{IEEEkeywords}

\section{Introduction}
Massive multiple input multiple output (MIMO) technology holds significant promise for enhancing wireless communication capacity by capitalizing on increased degrees of freedom \cite{telatar1999capacity}. Researchers have expressed considerable interest in exploring its applications in next-generation wireless systems \cite{larsson2014massive}. However, to fully leverage the spatial multiplexing and array gains of (massive) MIMO, precise channel state information at the Transmitter (CSIT) is indispensable \cite{bogale2011weighted,Hassibi:03}. While time division duplexing (TDD) MIMO systems can utilize channel reciprocity for CSIT acquisition using uplink pilots, acquiring CSIT for frequency division duplexing (FDD) systems presents challenges due to significant overheads \cite{Hoydis:13,marzetta2010noncooperative}.

\begin{figure}[htbp]
    \centering
    \includegraphics[scale=.15,trim={0cm 0cm 0cm 0cm}]{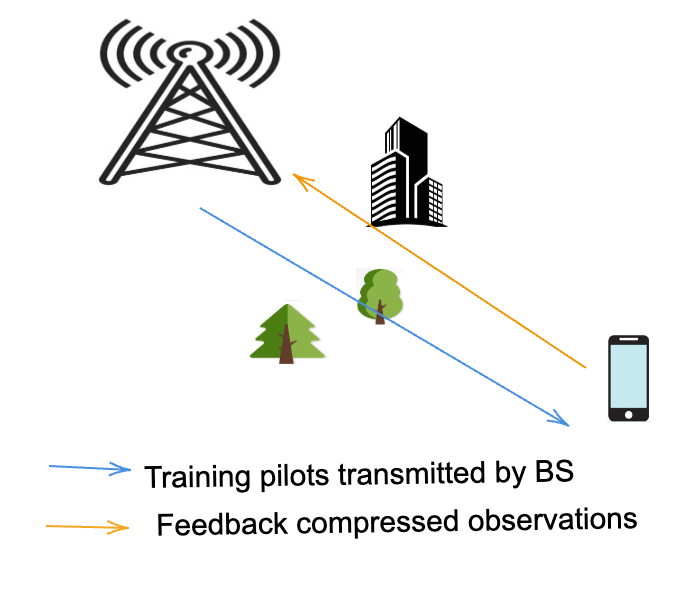}
        \caption{The BS transmits compressive training pilot to users. Each user locally obtains the compressive
measurement and feeds back to the BS. The BS then recovers the
CSIT based on the collected compressive measurements.}
    \label{fig:fdd_model}
\end{figure}

In FDD systems -- shown in Figure \ref{fig:fdd_model} -- CSIT at the base station (BS) is typically obtained by transmitting downlink pilot symbols and estimating the downlink channel state information locally at the user, which is then fed back to the BS via uplink signaling channels \cite{biguesh2006training}. Conventional methods, such as least squares (LS) and minimum mean square error (MMSE), for downlink CSI estimation require a large number of orthogonal pilot symbols, which is linearly proportional to the number of BS antennas \cite{biguesh2006training}. This overhead significantly increases in massive MIMO systems, rendering conventional methods in FDD systems impractical \cite{marzetta2010noncooperative}.

To tackle this challenge effectively, it is essential to leverage prior channel information in massive MIMO systems. The inherent angular sparsity observed in the user channel matrices within these systems \cite{zhou2007experimental} offers valuable insights that can be exploited through compressed sensing (CS) techniques \cite{daei2018improved,daei2019living,daei2019error}. However, existing CS-based methods in massive MIMO often assume an equal amount of information in both the angular and the spatial domains, or, alternatively, they assume the number of angular bins to be equal to the number of BS antennas, while indeed the angular domain is a much larger space than the spatial domain. This redundancy in angular domain can be effectively utilized through $\ell_1$ analysis optimization \cite{daei2018improved} rather than the conventional $\ell_1$ optimization used for capturing sparsity \cite{rao2014distributed,berger2010application,bajwa2010compressed}. By employing a larger number of angular bins than BS antennas, we can achieve better accuracy in sparse representation within the angular domain. 

Additionally, the dynamic nature of time-varying wireless channels provides a non-uniform prior distribution in the angular domain, which can be utilized in advance. For instance, due to the limited range of each user on the ground compared to a BS located at a higher elevation, only a fraction of the angular domain is active during user movement, while the rest remains inactive, leading to a non-uniform distribution in the angular domain. In this work, we propose a weighted $\ell_1$ analysis optimization that simultaneously captures both the redundant angular sparsity and the non-uniform angular prior distribution. We determine the optimal weights by minimizing the expected recovery error of the proposed optimization problem, which is equivalent to minimizing the statistical dimension \cite{daei2019living,daei2019error} of a certain cone. Note that the statistical dimension provides the minimum number of pilot symbols that the proposed optimization problem requires for exact channel estimation. Numerical experiments demonstrate the effectiveness of our approach in reducing the recovery error, thereby implying a reduction in the training and feedback overhead.

\textit{Notation}. $\mathds{P}$ and $\mathds{E}$ denote the probability and expectation operators, respectively. matrices and vectors are shown by bold upright letters. ${\rm cone}(S)$ denotes the conic hull of a set. $\partial f(\mx{x})$ denotes the sub-differential of the function $f$ at point $\mx{x}\in\mathbb{C}^n$. ${\rm erf}$ denotes the error function. $\mx{g}\in\mathbb{C}^n$ is a vector with i.i.d. elements and standard normal distribution. The sign of a vector is defined as ${\rm sgn}(\mx{x})\triangleq \tfrac{\mx{x}}{\|\mx{x}\|_2}$. $a\wedge b$ stands for the minimum of $a$ and $b$. $\mx{v}\in\mathbb{R}_{++}^{n}$ means that $v(i)>0 \forall i$. The infinity norm is defined as $\|\mx{z}\|_\infty=\max_i|z(i)|$. $x(l)$ is the $l$-th element of $\mx{x}\in\mathbb{C}^n$. The null space of a matrix is denoted by ${\rm null}(\cdot)$.

\section{System Model}\label{sec:system_model}
We consider a massive MIMO FDD system shown in Figure \ref{fig:fdd_model} with single-antenna users and assume the BS is equipped with $n$
antennas distributed as uniform linear array. For the downlink channel estimation in FDD system, the
BS transmits pilots to users. The users receive the pilots
and feeds the received signal back to the BS directly in an
ideal channel as in e.g., \cite{rao2014distributed,tseng2016enhanced}. The received downlink signal can be written as
\begin{align}\label{eq:main}
   \mx{y}=\mx{A}\mx{h}+\mx{e}\in\mathbb{C}^{m\times 1}, 
\end{align}
where $\mx{h}\in\mathbb{C}^{n\times 1}$ is the downlink spatial channel, $\mx{A}\in\mathbb{C}^{m\times n}$ is the downlink pilot matrix, $m\ll n$ is the pilot length and $\mx{e}\in\mathbb{C}^{m\times 1}$ is the additive measurement noise with $\|\mx{e}\|_2\le \eta $. Experimental studies have shown that the channel exhibits sparsity in the angular domain \cite{zhou2007experimental}. However, the dimension of the angular domain significantly exceeds that of the spatial domain, denoted as $n$. This suggests that upon applying an analysis operator $\bs{\Omega}\triangleq [\bs{\omega_1},...,\bs{\omega}_p]^T \in\mathbb{C}^{p\times n}, p\gg n$ to the spatial domain, we get to the angular sparse domain. Here, $\bs{\omega}_i=[1, ..., {\rm e}^{-2\pi (n-1)d ~\sin (\theta_i)}]^{\rm T}\in\mathbb{C}^{n\times 1}$ is the steering vector for angular bin $\theta_i$ and $d$ is the BS antenna spacing. Mathematically, we may express the downlink channel in the angular domain as
\begin{align}
 \mx{h}_{\rm a}=\bs{\Omega}\mx{h} \in\mathbb{C}^{p\times 1},
\end{align}
where the number of angular bins $p$ is much larger than the number $n$ of BS antennas. Here, $\mx{h}_{\rm a}\in\mathbb{C}^p$ is a sparse vector with analysis support $\mathcal{S}$ which represent angular bins that $h_{\rm a}(i)\neq 0, i=1,...,p$. The analysis Fourier operator $\bs{\Omega}$ is no longer orthonormal; however, it remains orthogonal. This implies that $\bs{\Omega}^{\rm H}\bs{\Omega}=\mx{I}_n$, but $\bs{\Omega}\bs{\Omega}^{\rm H}\neq\mx{I}_p$.  
The sparsity in the redundant angular domain is called analysis sparsity. To promote angular analysis sparsity, one can solve the following optimization problem \cite{daei2018improved}:
\begin{align}\label{eq.l1analysis}
 \mathsf{P}_{\bs{\Omega}}:~~&\min_{\mx{z}\in\mathbb{C}^n}\|\bs{\Omega z}\|_1\triangleq\sum_{i=1}^{p}|(\bs{\Omega z})_i| ~~{\rm s.t.}~\|\mx{y}_{m\times 1}-\mx{A}\mx{z}\|_2\le \eta,
 \end{align}
  In practice, due to the time-varying nature of the channel, there is a prior distribution of ${h}_{\rm a}(k)$ for each angular bin $k=1,...,p$ which together with redundant angular sparsity can be exploited. This angular distribution is non-uniformly distributed among angular bins. To promote angular sparsity and non-uniform structure together, we propose to use the following weighted optimization problem: 
  \begin{align}\label{eq.weightedl1}
 \mathsf{P}_{\bs{\Omega},\mx{v}}:~&\min_{\mx{z}\in\mathbb{C}^n}\|\bs{\Omega z}\|_{1,\mx{v}}\triangleq\sum_{i=1}^{p}v_i|(\bs{\Omega z})_i|~{\rm s.t.}\|\mx{y}_{m\times 1}-\mx{A}\mx{z}\|_2\le \eta,
 \end{align}
 where the weights $\mx{v}=\{v_1,\dots,v_p\}$ are some positive scalars served to penalize the angles that are less likely to be on the analysis support $\mathcal{S}$. In the next section, we provide an analytical way to determine the optimal weights in \eqref{eq.weightedl1}.
\section{Main result}\label{sec:main_res}
In this section, we aim to find an upper-bound for the recovery error of problem $\mathsf{P}_{\bs{\Omega},\mx{v}}$ in \eqref{eq.weightedl1}. Before that we require to define the expected joint sign parameter between $h_a(i)$ and $h_a(j)$ in the angular domain as follows:
\begin{align}
& \sigma_{i,j}\triangleq \mathds{E}[{\rm sgn}(h_a(i)){\rm sgn}(h_a(j))]=\nonumber\\
&\int\int {\rm sgn}(a){\rm sgn}(b)   f_{h_a(i),h_a(j)}(a,b) {\rm d}a {\rm d}b  
\end{align}
where $  f_{h_a(i),h_a(j)}(a,b)$ is the joint PDF of $h_a(i)$ and $h_a(j)$ for any $i,j=1,..., p$. The joint support distribution and the support distribution are also defined as $\beta_{i,j}\triangleq \mathds{P}(i\in\mathcal{S}, j\in\mathcal{S})$ and $\beta_i\triangleq \mathds{P}(i\in\mathcal{S})$ for any $i,j=1,..., p$.

\begin{thm}\label{thm.upper} 
    Let $\widehat{\mx{h}}\in\mathbb{C}^{n}$ be the solution of the optimization problem  $\mathsf{P}_{\bs{\Omega},\mx{v}}$.  Then, for any $\mx{v}\in\mathbb{R}_{+}^p$, the expected recovery error of $\mathsf{P}_{\bs{\Omega},\mx{v}}$ is bounded above by the following expression:
     \begin{align}\label{eq:error_bound}
         \mathds{E}_{\mx{h}}[\|\widehat{\mx{h}}-\mx{h}\|_2\le \tfrac{2\eta}{\max\{\sqrt{m-1}-\sqrt{\mathds{E}_{\mx{h}}[\delta(\mathcal{D}(\|\bs{\Omega}\cdot\|_{1,\mx{v}},\mx{h}))]}-a,0\}}+c
     \end{align}
    with probability at least $1-{\rm e}^{-\tfrac{a^2}{2}}$ in which $c>0$ is a constant and $\mathds{E}_{\mx{h}}[\delta(\mathcal{D}(\|\bs{\Omega} \cdot\|_{1,\mx{v}},\mx{h}))]$ is the expected required sample complexity of $\mathsf{P}_{\bs{\Omega},\mx{v}}$ which is bounded from above as follows:
     \begin{align}\label{eq.upper_expected}
 	&\mathds{E}_{\mx{h}}[\delta(\mathcal{D}(\|\bs{\Omega} \cdot\|_{1,\mx{v}},\mx{h}))]\le\nonumber\\
 	& \inf_{t\ge 0}\Big\{ n-\tfrac{\Big(\sum_{k=1}^p(1-\beta_k)\|\bs{\omega}_k\|_2{\rm erf}(\tfrac{tv_k}{\sqrt{2}})\Big)^2}{F(t,\mx{v})}\Big\},
 	% &F(t,\mx{v})=\sum_{k,k'=1}^p\Bigg[(tv_k)(tv_{k'})\bs{\omega}_k^{\rm H}\bs{\omega}_k\sigma_k\sigma_{k'}\beta_k\beta_{k'}+\tfrac{(\bs{\omega}_k^{\rm H}\bs{\omega}_k)^2}{\|\bs{\omega}_k\|_2\|\bs{\omega}_{k'}\|_2}\nonumber\\
 	% &\Big({\rm erf}(\tfrac{tv_{\rm min}}{\sqrt{2}})-q(tv_{\rm max})(tv_k)(tv_{k'})\Big)(1-\beta_k)(1-\beta_{k'})\Bigg],
  \end{align}
 	% \end{align}
 	where
  \begin{align}\label{eq:F_t_v}
    & F(t,\mx{v})\triangleq\sum_{i,j=1}^p (tv_i)(tv_j)\bs{\omega}_i^{\rm H}\bs{\omega}_j\sigma_{i,j}+\sum_{i,j=1}^p\tfrac{(\bs{\omega}_i^{\rm H}\bs{\omega}_j)^2}{\|\bs{\omega}_i\|_2\|\bs{\omega}_j\|_2}\nonumber\\
&\bigg[{\rm erf}(\tfrac{tv_{\min}}{\sqrt{2}})-q(tv_{\max})(tv_i)(tv_j)\bigg](1-\beta_i-\beta_j+\beta_{i,j})
 \end{align}
 and $q(t)\triangleq\sqrt{\tfrac{2}{\pi}}\tfrac{{\rm e}^{-\tfrac{t^2}{2}}}{t}+{\rm erf}(\tfrac{t}{\sqrt{2}})-1$.
\end{thm}
Proof. See Appendix \ref{proof.thm_upper}.

Our proposed strategy to find optimal weights is to minimize the expected recovery error provided in \eqref{eq:error_bound}. This leads to minimizing the upper-bound on the expected statistical dimension provided in \eqref{eq.upper_expected}. Specifically, the optimal weights can be found by solving the following optimization problem:
\begin{align}
  \mx{v}^\star=\mathop{\arg \min}_{\mx{v}>0}\inf_{t> 0}\Big\{ n-\tfrac{\Big(\sum_{k=1}^p(1-\beta_k)\|\bs{\omega}_k\|_2{\rm erf}(\tfrac{tv_k}{\sqrt{2}})\Big)^2}{F(t,\mx{v})}\Big\}.
 \end{align} 
The observant reader might have noticed that altering the multiplicative scaling of $\mx{v}$ does not impact the optimization $\mathsf{P}_{\Omega,\mx{v}}$ in \eqref{eq.weightedl1}, given that the weight vector's values hold significance relative  to each other. Hence, we can redefine $t\mx{v}$ simply as $\mx{v}$, treating it as a single variable and transform the latter optimization into
 \begin{align}\label{eq:optimal_weights}
\mx{v}^\star=\mathop{\arg \min}_{\mx{v}\in\mathbb{R}_{++}^p}
\Big\{\tfrac{2\eta}{\max\{\sqrt{m-1}-\sqrt{\overline{\delta}(\mx{v})}-a,0\}}\Big\}
\end{align}
where 
\begin{align}
\overline{\delta}(\mx{v})\triangleq  n-\tfrac{\Big(\sum_{k=1}^p(1-\beta_k)\|\bs{\omega}_k\|_2{\rm erf}(\tfrac{v_k}{\sqrt{2}})\Big)^2}{F_1(\mx{v})},
 \end{align}
 and 
 \begin{align}
 &F_1(\mx{v})\triangleq\sum_{k,k'=1}^p\Bigg[v_k v_{k'}\bs{\omega}_k^{\rm H}\bs{\omega}_{k'}\sigma_{k,k'}+\tfrac{(\bs{\omega}_k^{\rm H}\bs{\omega}_k)^2}{\|\bs{\omega}_k\|_2\|\bs{\omega}_{k'}\|_2}\nonumber\\
 &\Big({\rm erf}(\tfrac{v_{\rm min}}{\sqrt{2}})-q(v_{\rm max})v_k v_{k'}\Big)(1-\beta_k-\beta_{k'}+\beta_{k,k'})\Bigg].
 \end{align}

 \section{Simulations}\label{sec.simulation}
 \begin{figure}[t]
 	\centering
 	\includegraphics[scale=.3,trim={3cm 6cm 3cm 8cm}]{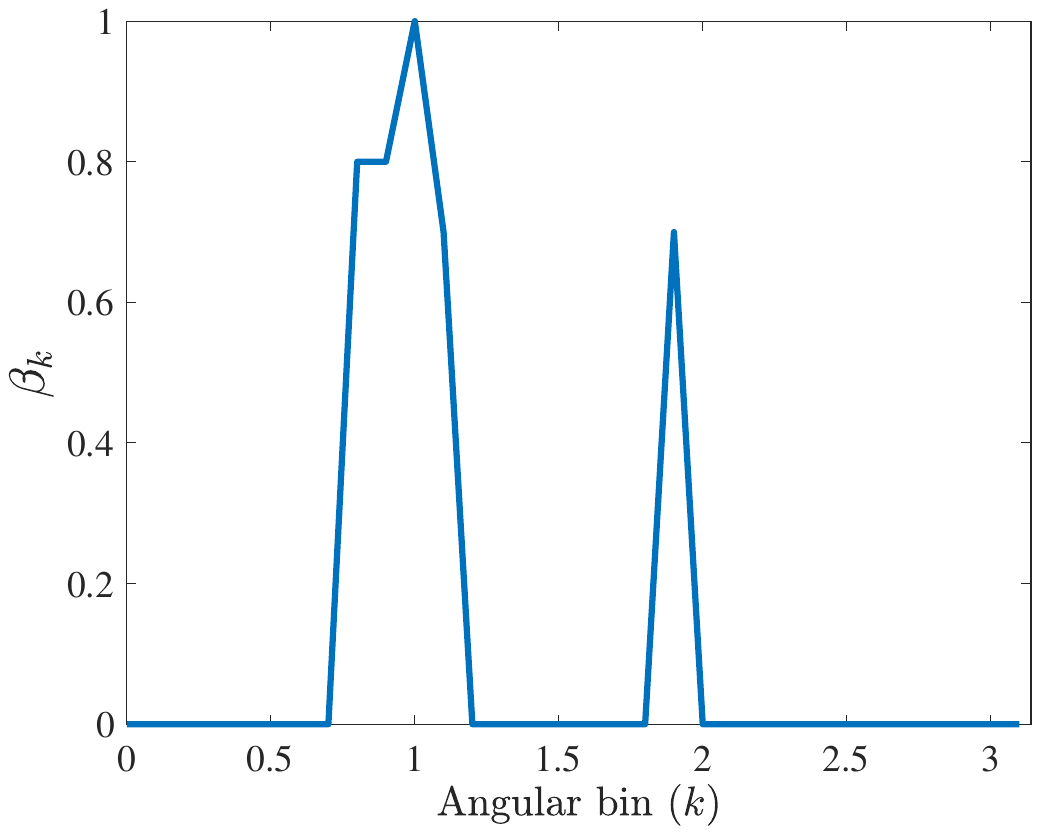}
 	\caption{The prior support distribution $\beta_k\triangleq\mathds{P}(k\in\mathcal{S}), k=1,..., p$ for a user considered in Section \ref{sec.simulation}.}\label{fig.support_dist}
 \end{figure}
 \begin{figure}[t]
 	\hspace{-0cm}
 	\includegraphics[scale=.13,trim={8cm 2cm 0cm 7cm},clip]{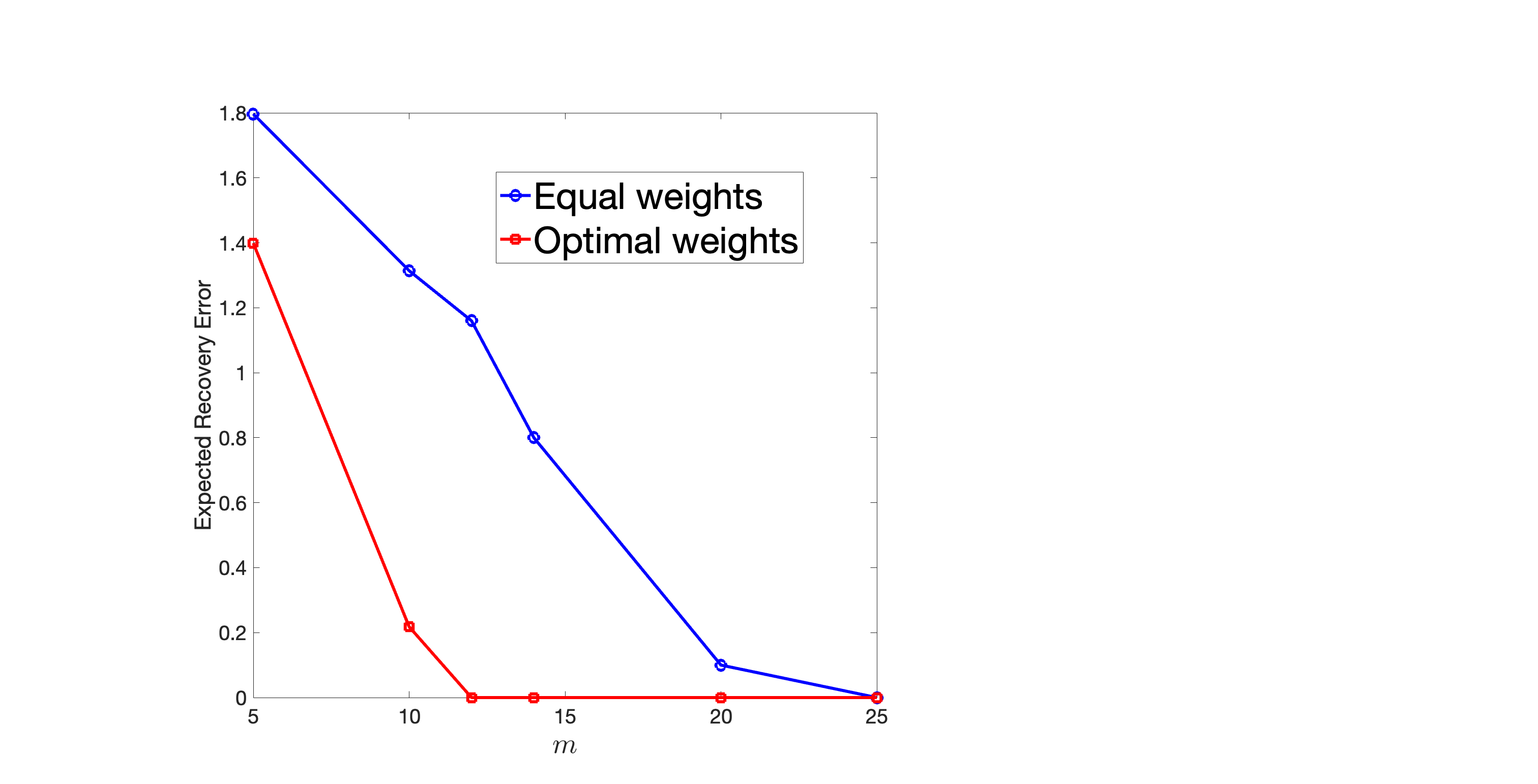}
 	\caption{This figure shows the expected recovery error versus the number of pilot symbols. The blue lines shows the recovery performance of $\mathsf{P}_{\bs{\Omega}}$ with constant weights  while the red line shows the performance of our proposed optimal weights provided in \eqref{eq:optimal_weights}.}\label{fig.recovery_error}
 \end{figure}
 In this section, we examine a numerical experiment to evaluate the performance of our method.
 % In the first experiment, we examine an example in which the signal of interest is random analysis-sparse with parameters $p=34, n=30$. To construct $\bs{\Omega}$, we follow the approach of \cite{kabanava2015analysis,nam2013cosparse} and first obtained orthonormal basis of the column span of a random matrix. Then, a non-tight frame $\bs{\Omega}$ is generated by changing the $\ell_2$ norms of rows. 
 The support $\mathcal{S}$ is randomly selected from the subset $\{1,...,p\}$ with known distribution $\beta_k, k=1,..., p$ shown in Figure \ref{fig.support_dist} and the spatial channel is generated as $\mx{h}={\rm null}(\bs{\Omega}_{\overline{\mathcal{S}}})\mx{c}$ where $\mx{c}$ is a random vector uniformly distributed on the uniform sphere and $\overline{\mathcal{S}}$ is the complement of $\mathcal{S}$ in the angular domain. This ensures that $\mx{h}_{\rm a}$ has an analysis support $\mathcal{S}$. We form the pilot matrix $\mx{A}_{m\times n}$ from an i.i.d. Gaussian distribution and perform the signal recovery using two methods: $\ell_1$ analysis, weighted $\ell_1$ analysis (Problem $\mathsf{P}_{\bs{\Omega},\mx{v}}$) with the optimal weights which are provided in \eqref{eq:optimal_weights}. The proposed near-optimal weights are obtained using support distribution $\{\beta_k\}_{k=1}^p$ and PDF of the channel in the angular domain i.e. $\{f_{h_a(k)}(\cdot)\}_{k=1}^p$ as prior information. Figure \ref{fig.recovery_error} shows the expected error defined by $\mathds{E}_{\mx{h},\mx{A}}\|\mx{h}-\widehat{\mx{h}}\|_2$ where $\widehat{\mx{h}}$ is the obtained estimate by solving $\mathsf{P}_{\bs{\Omega},\mx{v}}$. The latter expectation is calculated using $100$ independent Monte Carlo simulations over different $\mx{A}$ and $\mx{h}$. 
 As shown in Figure \ref{fig.recovery_error}, our proposed optimal weights provides a superior performance than the heuristic and constant weights in recovering the ground-truth channel from measurements $\mx{y}\in\mathbb{C}^m$.
\section{Conclusion}\label{sec:conclusion}
In this paper, we designed a new approach for downlink channel estimation in FDD massive MIMO systems in order to reduce the training and feedback overhead.
Specifically, we proposed a weighted optimization problem for channel estimation that simultaneously promote redundant angular sparsity and non-uniform angular distribution. We obtained a close-form expression for the expected recovery error of the proposed optimization problem dependent on the weights and the channel angular characteristics. We then found the optimal weights by minimizing this expression. Numerical results showed how the angular sparsity and non-uniform angular distribution contributes to enhancing the CSIT estimation quality in massive MIMO systems.

\section*{Appendix A: Proof of Theorem \ref{thm.upper}}\label{proof.thm_upper}
The recovery error in the problem $\mathsf{P}_{\bs{\Omega}}$ provided in \eqref{eq.l1analysis} is explicitly connected to the statistical dimension of the descent cone produced by the objective function $\|\bs{\Omega}\cdot\|_{1,\mx{v}}$ at the ground-truth channel $\mx{h}$. This is stated in the following bound \cite[Corrollary 3.5]{tropp2015convex}, \cite{daei2018improved}:
\begin{align}\label{eq:err_upper1}
 \|\widehat{\mx{h}}-\mx{h}\|_2\le \tfrac{2\eta}{\max\{\sqrt{m-1}-\sqrt{\delta(\mathcal{D}(\|\bs{\Omega}\cdot\|_{1,\mx{v}},\mx{h}))}-a,0\}},
\end{align}
in which $\mathsf{P}_{\bs{\Omega},\mx{v}}$ fails with probability at most ${\rm e}^{-\tfrac{a^2}{2}}$. 
By taking the expectation from both sides of \eqref{eq:err_upper1} and using the following Jensen's gap inequality \cite[Equation 2.1]{jensen_gap} and \cite{jensen}:
\begin{align}
 \mathds{E}[\tfrac{1}{\max\{b-\sqrt{x},0\}}]<\tfrac{1}{\max\{b-\sqrt{\mathds{E}[x]},0\}}+c,   
\end{align}
 for any $x<b^2$ with $b>0$, it holds that
\begin{align}
 \scalebox{.95}{$\mathds{E}_{\mx{h}}[\|\widehat{\mx{h}}-\mx{h}\|_2]\le 
 \tfrac{2\eta}{\max\{\sqrt{m-1}-\sqrt{\mathds{E}_{\mx{h}}[\delta(\mathcal{D}(\|\bs{\Omega}\cdot\|_{1,\mx{v}},\mx{h}))]}-a,0\}}  +c,  $}
\end{align}
where $c>0$ is an upper-bound for the variance of the statistical dimension. Now, finding an upper-bound for the error requires to find an upper-bound for the expected statistical dimension.
First, we begin with the definition of statistical dimension, i.e.
\begin{align}\label{eq.rel1}
&\delta(\mathcal{D}(\|\bs{\Omega}\cdot\|_{1,\mx{v}},\mx{h}))= \mathds{E}_{\mx{g}}{\rm dist}^2(\mx{g},{\rm cone}\partial \|\bs{\Omega}\cdot\|_1(\mx{h}))=\nonumber\\
&\mathds{E}_{\mx{g}} \inf_{t\ge 0}\inf_{\mx{z}\in\partial \|\bs{\Omega} \cdot\|_{1,\mx{v}}({\mx{h}})}\|\mx{g}-t\mx{z}\|_2^2.
\end{align}
By passing $\mathds{E}_{\mx{g}}$ through the first infimum, we reach an upper-bound as follows:
\begin{align}\label{eq.rel2}
&\mathds{E}_{\mx{g}}\inf_{t\ge 0}\inf_{\mx{z}\in\partial \|\bs{\Omega} \cdot\|_1({\mx{h}})}\|\mx{g}-t\mx{z}\|_2^2\le \inf_{t\ge 0}\mathds{E}_{\mx{g}}
\inf_{\mx{z}\in\partial \|\bs{\Omega} \cdot\|_{1,\mx{v}}({\mx{h}})}\|\mx{g}-t\mx{z}\|_2^2\nonumber\\
&=\inf_{t\ge 0}\mathds{E}_{\mx{g}}
\inf_{\|\mx{z}\|_{\infty}\le 1}\|\mx{g}-t\bs{\Omega}^{\rm H}\big(\mx{v}\odot{\rm sgn}(\bs{\Omega}\mx{h})\big)-t\bs{\Omega}^{\rm H}\big(\mx{v}\odot\mx{z}\big)\|_2^2\nonumber\\
&\triangleq B_u,
\end{align}
where in the last expression above, we used the chain rule of subdifferential (see \cite[Chapter 23]{rockafellar1970convex}) i.e. ${\partial\|\bs{\Omega}\cdot\|_{1,\mx{v}}(\mx{h})=\bs{\Omega}^{\rm H}\partial \|\cdot\|_{1,\mx{v}}(\bs{\Omega}\mx{h})}$ and the well-established relation for subdifferential of norm functions (see for example \cite[Proposition 1]{daei2019exploiting}) i.e.
\begin{align}
\partial \|\cdot\|_{1,\mx{v}}(\bs{\Omega}\mx{h})=\{\mx{v}\odot\big({\rm sgn}(\bs{\Omega}\mx{h})+\mx{z}\big): ~\|\mx{z}\|_{\infty}\le 1\}.
\end{align}
In \cite{daei2019error}, it is shown that $B_u$ is tight for different kinds of analysis operators. The expression $B_u$ does not seem to have a closed-form formula in general. Hence, we seek for an upper-bound of $B_u$. For this purpose, we substitute $\mx{z}$ with a special $\mx{z}'=[z_1',..., z_p']^p$ whose elements are defined as 
\begin{align}\label{eq.choice}
{z}_i'\triangleq\left\{\begin{array}{lr}
0,&i\in\mathcal{S}\\
{\rm sgn}(\bs{\omega}_i^{\rm H}\mx{g})\big(\tfrac{(tv_i)^{-1}\lambda |\bs{\omega}_i^{\rm H}\mx{g}|}{\|\bs{\omega}_i\|_2}\wedge 1\big),&i\in\overline{\mathcal{S}}
\end{array}\right\},
\end{align}
where $\lambda>0$ is a flexible tuning parameter. With this choice of $\mx{z}$, we can proceed with \eqref{eq.rel2} as follows:
\begin{align}\label{eq.rel4}
&\mathds{E}_{\mx{g}}
\|\mx{g}-t\bs{\Omega}^{\rm H}\big(\mx{v}\odot{\rm sgn}(\bs{\Omega}\mx{h})\big)-t\bs{\Omega}^{\rm H}\big(\mx{z}'\odot \mx{v}\big)\|_2^2=\nonumber\\
&\mathds{E}_{\mx{g}}\|\mx{g}\|_2^2-
2t\underbrace{\mathds{E}_{\mx{g}}\langle \mx{g},\bs{\Omega}^{\rm H}\big(\mx{v}\odot{\rm sgn}(\bs{\Omega}\mx{h})\big)\rangle}_{0}\nonumber\\
&-\underbrace{2t\mathds{E}_{\mx{g}}\langle \mx{g}, \bs{\Omega}^{\rm H}\big(\mx{v}\odot\mx{z}'\big)\rangle}_{\triangleq V_1}+\underbrace{t^2\mathds{E}_{\mx{g}}\|\bs{\Omega}^{\rm H}\big(\mx{v}\odot\mx{z}'\big)\|_2^2}_{\triangleq V_2}+\nonumber\\
&\underbrace{t^2\|\bs{\Omega}^{\rm H}\big(\mx{v}\odot{\rm sgn}(\bs{\Omega x})\big)\|_2^2}_{\triangleq V_3}=n-V_1+V_2+V_3.
\end{align}
Now, we calculate each term above one by one. First, consider
\begin{align}
&V_1=2t\mathds{E}_{\mx{g}}\langle \bs{\Omega}\mx{g},\mx{v}\odot\mx{z}'\rangle=2t\sum_{i\in\overline{\mathcal{S}}}v_i\mathds{E}_{\bm g}\big[\bs{\omega}_i^{\rm H}\mx{g}z_i'\big]\nonumber\\
&
=2\sum_{i\in\overline{\mathcal{S}}}\mathds{E}_{\bm g}\bigg[|\bs{\omega}_i^{\rm H}\mx{g}|\big(\tfrac{\lambda |\bs{\omega}_i^{\rm H}\mx{g}|}{\|\bs{\omega}_i\|_2}\wedge tv_i\big)\bigg],
\end{align}
where we used the relation \eqref{eq.choice} in the last expression. Since $\bs{\omega}_i^{\rm H}\mx{g}\sim \mathcal{N}(\mx{0},\|\bs{\omega}_i\|_2^2)$, it is straightforward to obtain the expectation inside the summation and verify that
\begin{align}
V_1=2\sum_{i\in\overline{\mathcal{S}}}\|\bs{\omega}_i\|_2\mathds{E}_{\mx{g}}\big[|\mx{g}|(\lambda |\mx{g}|\wedge tv_i)\big].
\end{align}
We proceed by calculating
\begin{align}
&\mathds{E}_{\mx{g}}\big[|\mx{g}|(\lambda |\mx{g}|\wedge tv_i)\big]=\int_{0}^{\tfrac{tv_i}{\lambda}}\lambda\sqrt{\tfrac{2}{\pi}}z^2 {\rm e}^{-\tfrac{z^2}{2}}{\rm d}z+\nonumber\\
&\int_{\tfrac{tv_i}{\lambda}}^{\infty}\sqrt{\tfrac{2}{\pi}}t v_i z {\rm e}^{-\tfrac{z^2}{2}}{\rm d}z=\sqrt{\tfrac{2}{\pi}} t v_i {\rm e}^{-\tfrac{t^2v_i^2}{2\lambda^2}}-\sqrt{\tfrac{2}{\pi}} t v_i {\rm e}^{-\tfrac{t^2v_i^2}{2\lambda^2}}+\nonumber\\
&\lambda {\rm erf}(\tfrac{tv_i}{\sqrt{2}\lambda}),
\end{align}
which leads to
\begin{align}\label{eq.V1relfinal}
V_1=2\sum_{i\in\overline{\mathcal{S}}}\lambda \|\bs{\omega}_i\|_2 {\rm erf}(\tfrac{tv_i}{\sqrt{2}\lambda}).
\end{align}
For $V_2$, we have
\begin{align}\label{eq.V2}
&V_2=t^2\mathds{E}_{\mx{g}}\langle \mx{v}\odot\mx{z}', \bs{\Omega}\bs{\Omega}^{\rm H} \big(\mx{v}\odot\mx{z}'\big)\rangle=t^2\sum_{i,j\in\overline{\mathcal{S}}}v_iv_j\bs{\omega}_i^{\rm H}\bs{\omega}_j\nonumber\\
&\mathds{E}_{\mx{g}}[z_i'z_j']=\sum_{i,j\in\overline{\mathcal{S}}}
(tv_i)(tv_j)\bs{\omega}_i^{\rm H}\bs{\omega}_j\mathds{E}_{\mx{g}}\bigg[{\rm sgn}(\bs{\omega}_i^{\rm H}\mx{g}){\rm sgn}(\bs{\omega}_j^{\rm T}\mx{g})\nonumber\\
&\big(\tfrac{(tv_i)^{-1}\lambda \bs{\omega}_i^{\rm H}\mx{g}}{\|\bs{\omega}_i\|_2}\wedge 1\big)\big(\tfrac{(tv_j)^{-1}\lambda \bs{\omega}_j^{\rm T}\mx{g}}{\|\bs{\omega}_j\|_2}\wedge 1\big)\bigg]=\sum_{i,j\in\overline{\mathcal{S}}}\lambda^2\bs{\omega}_i^{\rm H}\bs{\omega}_j\nonumber\\
&\mathds{E}_{\mx{g}}\bigg[{\rm sgn}(\bs{\omega}_i^{\rm H}\mx{g}){\rm sgn}(\bs{\omega}_j^{\rm T}\mx{g})\big(\tfrac{|\bs{\omega}_i^{\rm H}\mx{g}|}{\|\bs{\omega}_i\|_2}\wedge \tfrac{tv_i}{\lambda}\big)\big(\tfrac{|\bs{\omega}_j^{\rm T}\mx{g}|}{\|\bs{\omega}_j\|_2}\wedge \tfrac{tv_j}{\lambda}\big)\bigg]
\end{align}

Deriving a closed-form formula for the final expression above appears to be challenging in general. In such cases, we resort to an upper bound provided in the following lemma, adapted from \cite[Lemma 6.12]{genzel2017ell}.
\begin{lem}\cite[Lemma 6.12]{genzel2017ell}
Let $\mx{g}\sim \mathcal{N}(\mx{0}, \mx{I}_n)$. For any $\mx{z}_1, \mx{z}_2\in\mathbb{S}^{n-1}$, $\alpha_1, \alpha_2\ge 0$, and $\alpha_{\min}=\min\{\alpha_1,\alpha_2\}$, $\alpha_{\max}=\max\{\alpha_1,\alpha_2\}$, we have
\begin{align}
&\mathds{E}\bigg[{\rm sgn}(\mx{z}_1^{\rm T}\mx{g}){\rm sgn}(\mx{z}_2^{\rm T}\mx{g})\big(|\mx{z}_1^{\rm T}\mx{g}|\wedge \alpha_1\big)\big(|\mx{z}_2^{\rm T}\mx{g}|\wedge \alpha_2\big)\bigg]\le \nonumber\\
&|\langle \mx{z}_1, \mx{z}_2 \rangle|\bigg[ {\rm erf}(\tfrac{\alpha_{\min}}{\sqrt{2}})-q(\alpha_{\max})\alpha_1\alpha_2\bigg].
\end{align}
\end{lem}
By benefiting this lemma, we proceed \eqref{eq.V2} as follows:
 \begin{align}\label{eq.V2relfinal}
 V_2\le \lambda^2\sum_{i,j\in\overline{\mathcal{S}}}\tfrac{(\bs{\omega}_i^{\rm H}\bs{\omega}_j)^2}{\|\bs{\omega}_i\|_2\|\bs{\omega}_j\|_2}\bigg[{\rm erf}(\tfrac{tv_{\min}}{\lambda\sqrt{2}})-q(\tfrac{tv_{\max}}{\lambda})\tfrac{(tv_1)(tv_2)}{\lambda^2}\bigg].
 \end{align}
 Lastly, for $V_3$, it holds that
 \begin{align}\label{eq.V3relfinal}
 &V_3=t^2\langle \mx{v}\odot {\rm sgn}(\bs{\Omega}\mx{h}), \bs{\Omega}\bs{\Omega}^{\rm H} \big(\mx{v}\odot {\rm sgn}(\bs{\Omega}\mx{h})\big)\rangle=\nonumber\\
 &\sum_{i,j\in\mathcal{S}}(tv_i)(tv_j)\bs{\omega}_i^{\rm H}\bs{\omega}_j{\rm sgn}(\bs{\omega}_i^{\rm H}\mx{h}){\rm sgn}(\bs{\omega}_j^{\rm T}\mx{h}).
 \end{align}
 Combining \eqref{eq.V1relfinal}, \eqref{eq.V2relfinal}, \eqref{eq.V3relfinal}, and \eqref{eq.rel4}, leads to the following upper-bound for $\delta(\mathcal{D}(\|\bs{\Omega}\cdot\|_{1,\mx{v}},\mx{h}))$.
 \begin{align}
 &\delta(\mathcal{D}(\|\bs{\Omega}\cdot\|_{1,\mx{v}},\mx{h}))\le \inf_{t> 0}\inf_{\lambda> 0}\Bigg\{n+\sum_{i,j\in\mathcal{S}}(tv_i)(tv_j)\bs{\omega}_i^{\rm H}\bs{\omega}_j\nonumber\\
 &{\rm sgn}(\bs{\omega}_i^{\rm H}\mx{h}){\rm sgn}(\bs{\omega}_j^{\rm T}\mx{h})-2\sum_{i\in\overline{\mathcal{S}}}\lambda \|\bs{\omega}_i\|_2 {\rm erf}(\tfrac{tv_i}{\sqrt{2}\lambda})+\nonumber\\
 &\lambda^2\sum_{i,j\in\overline{\mathcal{S}}}\tfrac{(\bs{\omega}_i^{\rm H}\bs{\omega}_j)^2}{\|\bs{\omega}_i\|_2\|\bs{\omega}_j\|_2}\bigg[{\rm erf}(\tfrac{tv_{\min}}{\lambda\sqrt{2}})-q(\tfrac{tv_{\max}}{\lambda})\tfrac{(tv_1)(tv_2)}{\lambda^2}\bigg]\Bigg\}.
 \end{align}
 By substituting $t$ by $\lambda t$, we have
\begin{align}
&\delta(\mathcal{D}(\|\bs{\Omega}\cdot\|_{1,\mx{v}},\mx{h}))\le \inf_{t> 0}\inf_{\lambda> 0}\Bigg\{n+\lambda^2\sum_{i,j\in\mathcal{S}}(tv_i)(tv_j)\bs{\omega}_i^{\rm H}\bs{\omega}_j\nonumber\\
&{\rm sgn}(\bs{\omega}_i^{\rm H}\mx{h}){\rm sgn}(\bs{\omega}_j^{\rm T}\mx{h})-2\sum_{i\in\overline{\mathcal{S}}}\lambda \|\bs{\omega}_i\|_2 {\rm erf}(\tfrac{tv_i}{\sqrt{2}})+\nonumber\\
&\lambda^2\sum_{i,j\in\overline{\mathcal{S}}}\tfrac{(\bs{\omega}_i^{\rm H}\bs{\omega}_j)^2}{\|\bs{\omega}_i\|_2\|\bs{\omega}_j\|_2}\bigg[{\rm erf}(\tfrac{tv_{\min}}{\sqrt{2}})-q(tv_{\max})(tv_1)(tv_2)\bigg]\Bigg\}.
\end{align}

By taking expectation with respect to $\mx{h}$ from the latter expression and simplifying, it follows that
% \begin{align}
% &\mathds{E}_{\mx{h}}\delta(\mathcal{D}(\|\bs{\Omega} \cdot\|_{1,\mx{v}},\mx{h}))\le \inf_{t> 0}\inf_{\lambda> 0}\Bigg\{n+\lambda^2\sum_{i,j=1}^p (tv_i)(tv_j)\nonumber\\
% &\bs{\omega}_i^{\rm H}\bs{\omega}_j\mathds{E}_{\mx{h}}\big[{\rm sgn}(\bs{\omega}_i^{\rm H}\mx{h}){\rm sgn}(\bs{\omega}_j^{\rm T}\mx{h})\big]\nonumber\\
% &-2\sum_{i=1}^p\lambda \|\bs{\omega}_i\|_2 {\rm erf}(\tfrac{tv_i}{\sqrt{2}})\mathds{E}_{\mx{h}}\big[1_{i\in\overline{\mathcal{S}}}\big]+\lambda^2\sum_{i,j=1}^p\tfrac{(\bs{\omega}_i^{\rm H}\bs{\omega}_j)^2}{\|\bs{\omega}_i\|_2\|\bs{\omega}_j\|_2}\nonumber\\
% &\bigg[{\rm erf}(\tfrac{tv_{\min}}{\sqrt{2}})-q(tv_{\max})(tv_i)(tv_j)\bigg]\mathds{E}_{\mx{h}}\big[1_{i\in\overline{\mathcal{S}}}1_{j\in\overline{\mathcal{S}}}\big]\Bigg\}.
% \end{align}
% By further simplifying, we can reach
\begin{align}\label{eq.relll1}
&\scalebox{.9}{$\mathds{E}_{\mx{h}}\delta(\mathcal{D}(\|\bs{\Omega} \cdot\|_{1,\mx{v}},\mx{h}))\le \inf_{\substack{t> 0\\
\lambda>0}}\Bigg\{n+\lambda^2\sum_{i,j=1}^p (tv_i)(tv_j)\bs{\omega}_i^{\rm H}\bs{\omega}_j\sigma_{i,j}$}\nonumber\\
&-2\sum_{i=1}^p\lambda \|\bs{\omega}_i\|_2 {\rm erf}(\tfrac{tv_i}{\sqrt{2}})(1-\beta_i)+\lambda^2\sum_{i,j=1}^p\tfrac{(\bs{\omega}_i^{\rm H}\bs{\omega}_j)^2}{\|\bs{\omega}_i\|_2\|\bs{\omega}_j\|_2}\nonumber\\
&\bigg[{\rm erf}(\tfrac{tv_{\min}}{\sqrt{2}})-q(tv_{\max})(tv_i)(tv_j)\bigg](1-\beta_i-\beta_j+\beta_{i,j}))\Bigg\}.
\end{align}

Now, by taking derivative of the term inside the infimum with respect to $\lambda>0$ and set it to zero, we have: $\lambda^\star=\tfrac{\sum_{i=1}^p \|\bs{\omega}_i\|_2 {\rm erf}(\tfrac{tv_i}{\sqrt{2}})(1-\beta_i)}
 {F(t,v)}$,
 where $F(t,v)$ is defined in \eqref{eq:F_t_v}.
 By replacing $\lambda^{\star}$ into \eqref{eq.relll1}, we have:
 \begin{align}\label{eq:sample_upper}
   &\scalebox{.9}{$  \mathds{E}_{\mx{h}}\delta(\mathcal{D}(\|\bs{\Omega} \cdot\|_{1,\mx{v}},\mx{h}))\le \inf_{t>0}
    n-\tfrac{\Big(\sum_{k=1}^p(1-\beta_k)\|\bs{\omega}_k\|_2{\rm erf}(\tfrac{tv_k}{\sqrt{2}})\Big)^2}{F(t,\mx{v})}$}
 \end{align}
 Combining \eqref{eq:err_upper1} and \eqref{eq:sample_upper} leads to the final result.

\bibliographystyle{IEEEtran}
\bibliography{references}

\end{document}